\documentclass[fleqn,10pt]{wlscirep}

\usepackage{amsmath,amssymb,graphicx,color}

\newcommand{\opr}{\omega_{\rm pr}}
\newcommand{\op}{\omega_{\rm p}}
\newcommand{\veps}{\varepsilon}

\title{Optically induced metal-to-dielectric transition in Epsilon-Near-Zero metamaterials}

\author[1]{R. M. Kaipurath}
\author[2]{M. Pietrzyk}
\author[1]{L. Caspani}
\author[1]{T. Roger}
\author[1,3]{M. Clerici}
\author[4,5]{C. Rizza}
\author[5]{A. Ciattoni}
\author[2,*]{A. Di Falco}
\author[1,$\dag$]{D. Faccio}
\affil[1]{School of Engineering and Physical Sciences, SUPA, Institute of Photonics and Quantum Sciences
Heriot-Watt University, Edinburgh EH14 4AS, UK}
\affil[2]{SUPA, School of Physics and Astronomy, University of  St. Andrews, St. Andrews KY16 9SS, UK}
\affil[3]{School of Engineering, University of Glasgow, Glasgow G12 8LT, UK}
\affil[4]{Dipartimento di Scienza e Alta Tecnologia, Universit\`a dell'Insubria, Via Valleggio 11, 22100 Como, Italy}
\affil[5]{Consiglio Nazionale delle Ricerche, CNR-SPIN, Via Vetoio 10, 67100 L'Aquila, Italy}

\affil[*]{adf10@st-andrews.ac.uk}
\affil[$\dag$]{d.faccio@hw.ac.uk}

\begin{abstract}
Epsilon-Near-Zero  materials exhibit a transition in the real part of the  dielectric permittivity from positive to negative value as a function of wavelength. Here we study metal-dielectric layered metamaterials in the homogenised regime (each layer has strongly subwavelength thickness) with zero real part of the permittivity in the near-infrared region. By optically pumping the metamaterial we experimentally show that close to the Epsilon-Equal-to-Zero (EEZ) wavelength the permittivity exhibits a marked transition from metallic (negative permittivity) to dielectric (positive permittivity) as a function of the optical power. Remarkably, this transition is linear as a function of pump power and occurs on time scales of the order of the 100 fs pump pulse that need not be tuned to a specific wavelength. The linearity of the permittivity increase allows us to express the response of the metamaterial in terms of a standard third order optical nonlinearity: this shows a clear inversion of the roles of the real and imaginary parts in crossing the EEZ wavelength, further supporting an optically induced change in the physical behaviour of the metamaterial.
\end{abstract}
\begin{document}

\flushbottom
\maketitle

\thispagestyle{empty}

\section*{Introduction}

Recent advances in metamaterial science have opened routes to unprecedented control over the optical properties of matter, with a wide array of applications and implications for novel light-matter interactions. Examples are the demonstration of negative index materials and, more recently, significant attention has been devoted to the behaviour of light in a medium with zero dielectric permittivity. We will refer to these materials as Epsilon-Equal-to-Zero (EEZ) materials with the implicit assumption that in all passive materials, the EEZ condition will only be met for the real part of the permittivity, $\varepsilon '$, (as a result of absorption that will always imply that the imaginary part is greater than zero) and at one single wavelength (due to dispersion). Such EEZ materials may either occur naturally at the plasma frequency or may result from engineering the propagation medium, for example so that light propagates in a waveguide near the cutoff frequency. Another option, investigated here, is to create a metamaterial (MM) made of deeply subwavelength alternating layers of dielectric and metal with thicknesses that are chosen such that $\varepsilon '$ is zero at a chosen wavelength. 
The linear properties of EEZ, often referred to as Epsilon-Near-Zero (ENZ) metamaterials  have been investigated in depth with a range of applications for example in novel waveguiding regimes  and for controlling the radiation pattern of electromagnetic sources \cite{alu1,alu2,eps0_1,eps0_2,eps0_3,eps0_4,eps0_5,eps0_6,eps0_7,eps0_8,eps0_9,eps0_11,eps0_12,eps0_15,eps0_14}.	
The ENZ condition  has also been predicted to have far-reaching consequences in terms of the effective optical nonlinearity of the metamaterial, but with limited experimental evidences \cite{eps0_13,ciattoni:2010,ciattoni2,ciattoni3,Ciattoni:10,zayats,Rizza_11,Suchowski:2013fa,husakou}. A compelling experimental evidence of the role of ENZ properties affecting the optical nonlinearity is the recent demonstration of efficient third harmonic generation due to the enhancement of the pump electric field longitudinal component in a uniform film of Indium-Tin-Oxide (ITO) \cite{ITO}.\\
A different, yet related area of study, is the search for materials that can be optically controlled so as to exhibit a sharp and rapid transition from metallic to dielectric (or vice versa), thus implying a fundamental change in the material properties. Examples that have been investigated and observed in literature are optically induced phase transitions (for example in Vanadium Oxide and other compounds) \cite{VO2} or sudden increase in conductivity in glass when optically pumped with single cycle pulses, close to the breakdown damage threshold \cite{Krausz_nature1,Krausz_nature2}. A metal-to-dielectric transition has also been theoretically proposed in metallo-dielectric stacks, obtained by optically pumping close to the EEZ wavelength \cite{husakou} with relatively fast $\sim1$ ps response times.
Here we experimentally investigate the optical behaviour of MMs made of deeply subwavelength alternating layers of fused silica glass and silver with thicknesses that are chosen such that $\varepsilon '=0$ in the near-infrared region, as shown  in Fig.~\ref{fig1}. We show that by optically pumping the MM far from the EEZ wavelength, it is possible to induce a marked and rapid (on the same time scale of the 100 fs pump pulse) transition from metallic ($\varepsilon '<0$) to dielectric ($\varepsilon '>0$). By changing the layer thickness of the MM we can also control the wavelength at which this transition occurs. Remarkably we find that the metal-dielectric transition occurs linearly as a function of the pump intensity. We show that this fact allows to  describe the optical behaviour in terms of a standard third order nonlinear susceptibility. Our measurement technique provides the full complex amplitude across a wide spectral range centred at the EEZ wavelength.

\section*{Results}
We fabricated the metamaterial samples by electron beam deposition of alternating layers of Ag and SiO$_2$ on a thick (1 mm) SiO$_2$ substrate with a total of 10 layers. The Ag thickness in each layer is kept at 5 nm whilst the  SiO$_2$ thickness is the same in each layer and tuned to 80-70 nm in order to provide the ENZ condition around 820-890 nm, in the centre of our laser tuning region. To obtain smooth continuous layers of silver on silica below the standard percolation limit, we seeded the deposition of each metal layer with 0.7 nm of Germanium \cite{Shalaev_10}. Figures~\ref{fig1}(a) and (b) show a photograph of one of the samples and an SEM image of the multilayer structure, respectively. More details on the fabrication process are provided in the Methods section.
\subsection*{Linear response}
\begin{figure}[hbt!]
\centering
\includegraphics[width=8cm]{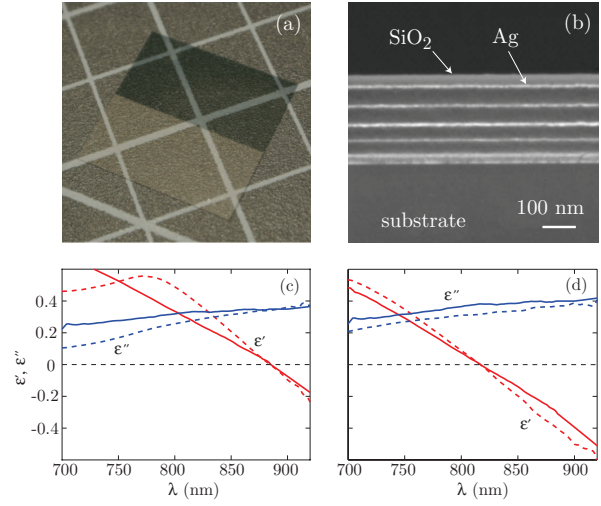}
\caption{ {\bf Pictures of the sample and linear characterisation.} (a) Photograph of the metamaterial sample, (b) SEM image of the metamaterial sample showing the alternating layers of metal (Ag, 5 nm)  and dielectric (SiO$_2$, 70 nm) with 35 nm top and bottom layers of SiO$_2$. The laser exciting the material nonlinear response is directed along the surface normal and is polarised parallel to the surface.  (c) The measured metamaterial linear dielectric permittivity (retrieved by fitting the experimental transmission and reflection spectra): real ($\varepsilon '$, red dashed curve) and imaginary ($\varepsilon ''$, blue dashed curve) parts.  The solid curves show the theoretical effective-medium predictions for $\varepsilon '$ and $\varepsilon ''$, as given by the formula $\varepsilon_{\textrm{eff}}=(l_{\rm m}\varepsilon_{\textrm{Ag}}+l_{\rm d}\varepsilon_{\textrm{SiO$_2$}})/(l_{\rm m}+l_{\rm d})$, where $l_{\rm m}$ is the thickness of the Ag layer and $l_{\rm d}$ is the thickness of the SiO$_2$ layer \cite{maxwell_garnett}. $\varepsilon '= 0$  for a wavelength of $\lambda=885$ nm. (d)  Same as in (c) with slightly tuned SiO$_2$ thickness so that the EEZ wavelength is shifted to 820 nm. \label{fig1}}
\end{figure}

The linear response (real and imaginary part of $\varepsilon$, $\varepsilon '$ and $\varepsilon ''$, respectively) was measured by a standard reflection/transmission measurement (see Methods for details) and is shown as the dashed lines in Figs.~\ref{fig1}(c) and (d) for two different samples  with SiO$_2$ thickness equal to 80 nm (referred to as ``sample A'' in the following) and 70 nm (``sample B''), respectively. $\varepsilon '$ is measured to be zero at 885 nm (sample A) and 820 nm (sample B).\\
Under the approximation of deeply subwavelength films, light polarised parallel to the film does not interact with each individual layer of the multilayer structure but rather with an effective homogenised medium whose complex dielectric susceptibility is given by $
\chi_{\rm eff}=(l_{\rm d}\chi_{\rm d}+l_{\rm m}\chi_{\rm m})/(l_{\rm d}+l_{ \rm m})$
where $\chi$ may represent either the linear susceptibility, $\chi^{(1)}$ or also the third order nonlinear susceptibility, $\chi^{(3)}$ \cite{sipe:1992}. $l$ indicates the thickness of the individual layer and the subscripts d and m refer to the dielectric and metal layers, respectively.\\
The solid lines in Figs.~\ref{fig1}(c) and (d) show the predictions for a homogenised material and are in excellent agreement with the measured data between 750 nm  and 920 nm, thus indicating that in this wavelength region the metamaterial is indeed behaving as an effectively uniform and homogenised medium.
\subsection*{Nonlinear response}
We measured the nonlinear response of the metamaterial by monitoring the changes in reflectivity, $\Delta R=R-R_{\rm lin}$, and transmissivity, $\Delta T=T-T_{\rm lin}$, in a pump and probe experiment (where $R_{\rm lin}$ and $T_{\rm lin}$ are the linear --without the pump-- reflectivity and transmissivity, respectively).
 The pump (with a fixed wavelength of 785 nm, pulse duration 100 fs, horizontally polarised) is at normal incidence on the sample and the probe (wavelength tuned in the 700-1000 nm region, 100 fs pulse duration, 50 Hz repetition rate, vertically polarised) is incident at a small $\sim1$ deg. angle with respect to the pump. The pump-probe delay was  adjusted to maximise the nonlinear effect i.e. was zero within the precision of the pulse duration. The probe power is always kept extremely low so that alone it does not induce any nonlinear effects whilst the pump power is varied between zero (pump blocked) and $\sim30$ GW/cm$^2$. 
 
\begin{figure}[bt]
\begin{center}
\includegraphics[width=12cm]{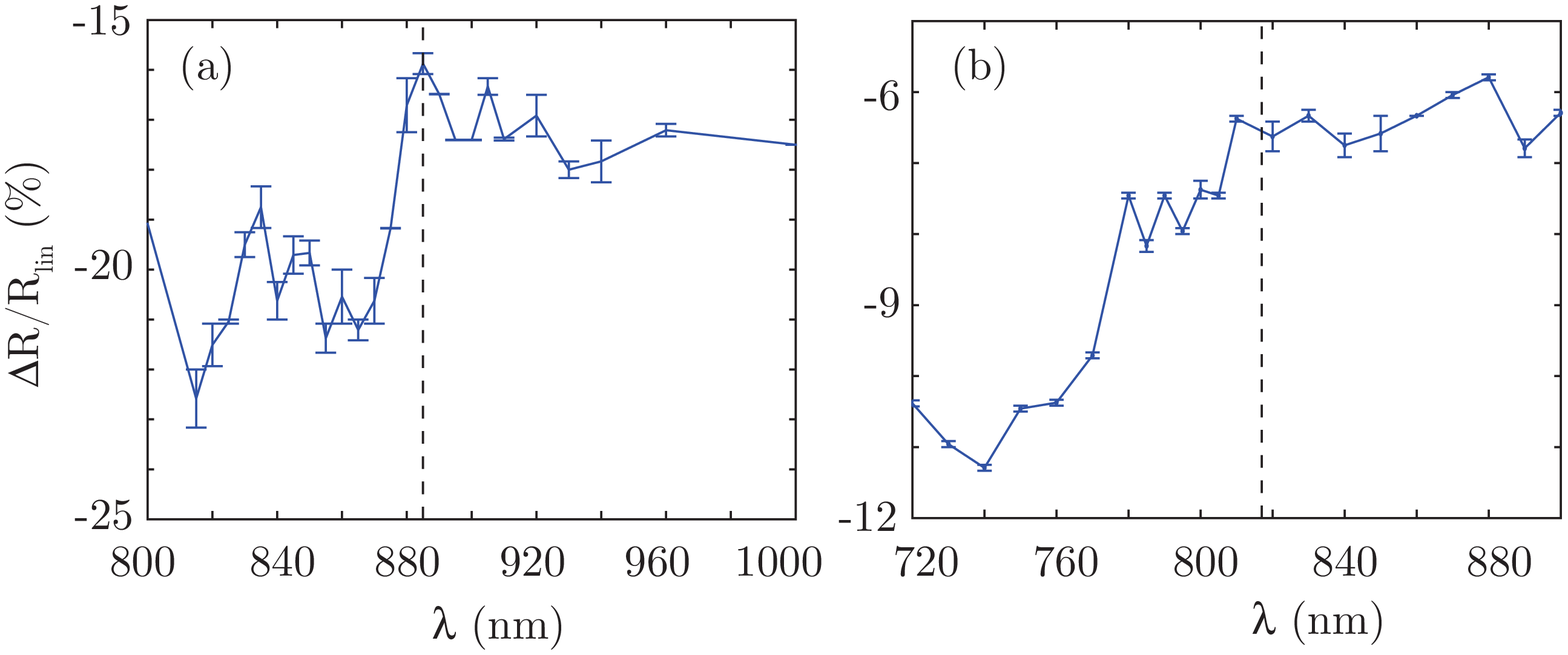}
\caption{{\bf Experimental results: nonlinear (pumped) reflectivity as a function of the probe wavelength.}  Pump and probe reflectivity variation for the two metamaterial samples [sample A (a) and sample B in (b)] around the EEZ wavelength (indicated with a vertical dashed line). The pump wavelength is kept fixed at 785 nm whilst the probe wavelength is varied (horizontal axis). The pump intensity is 20 GW/cm$^2$ and 10 GW/cm$^2$, for sample A and B, respectively. The data show a step-like behaviour in proximity to the EEZ wavelength. \label{fig2}}
\end{center}
\end{figure}

\begin{figure}[!t]
\begin{center}
\includegraphics[width=8.5cm]{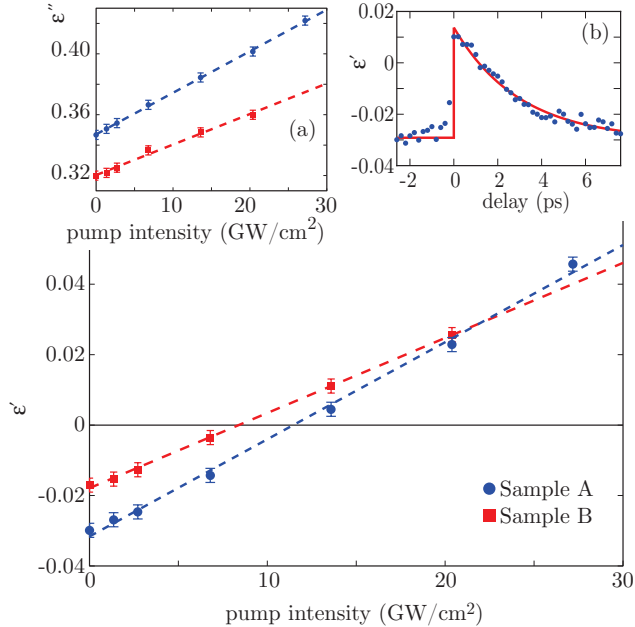}
\caption{{\bf Nonlinear (pumped) behaviour of the permittivity.} Real part of $\veps$ ($\veps '$) versus input pump intensity for sample A (blue circles) and B (red squares). The dashed lines show linear fits to the experimental data. Inset (a) shows instead the imaginary party of $\veps$ ($\veps ''$). Inset (b) shows the variation in $\veps '$ versus pump-probe delay for sample A measured at 17 GW/cm$^2$ pump intensity. $\lambda_{\rm pump}=$785 nm, $\lambda_{\rm probe}=$890 and 825 nm, for sample A and B, respectively. The error bars have been evaluated as standard deviation over 10 samples. \label{fig3a}}
\end{center}
\end{figure}

\begin{figure}[!t]
\begin{center}
\includegraphics[width=12cm]{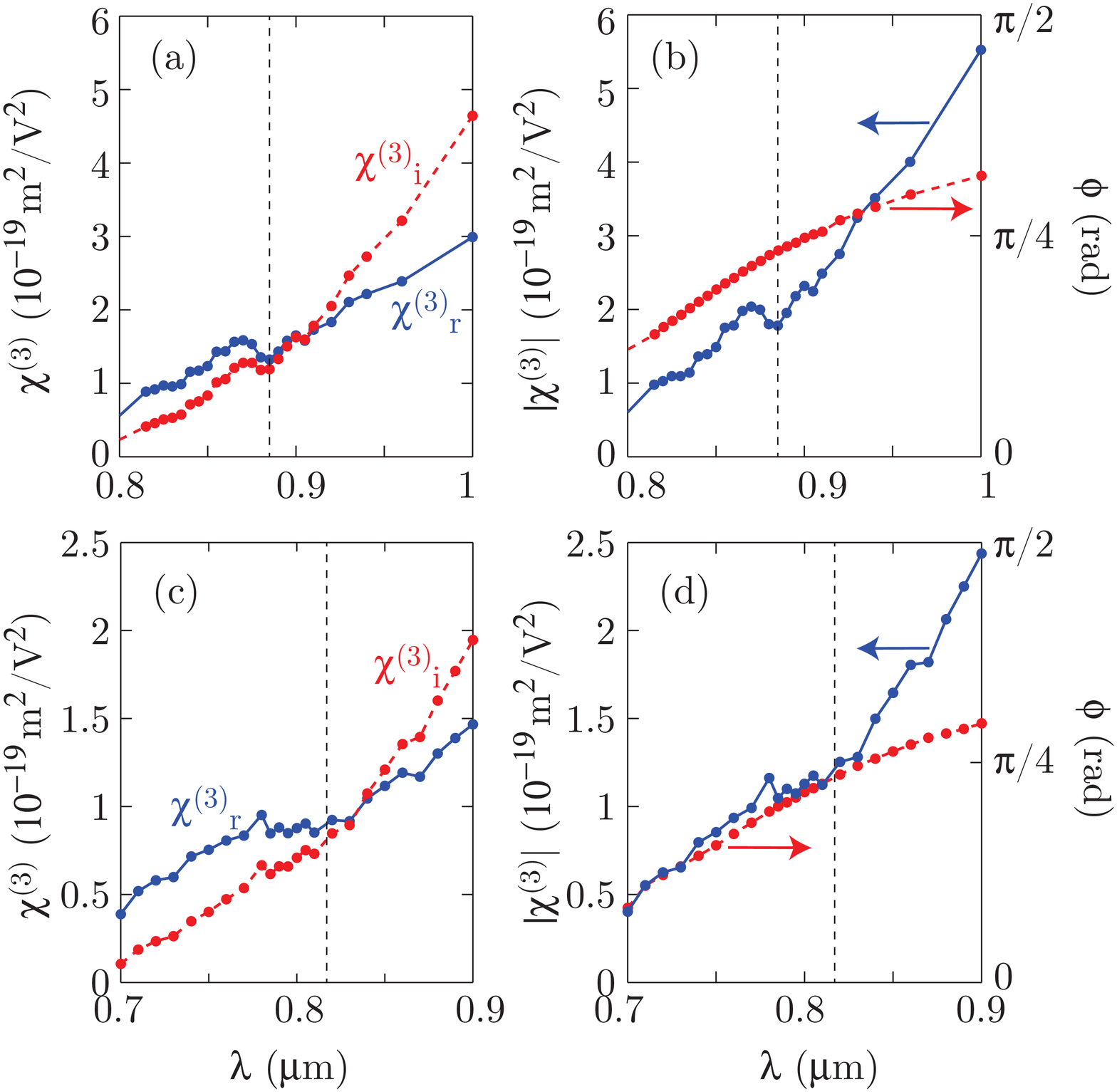}
\caption{{\bf Experimental values of the third-order nonlinear susceptibility.} Real and imaginary parts of the third order nonlinearity $\chi^{(3)}_r$ (solid blue line) and $\chi^{(3)}_i$ (dashed red line), respectively for sample A (a) and sample B (c). The same data is also plotted as the absolute value $|\chi^{(3)}|$  (solid blue line) and phase $\phi$ (dashed red line) for sample A in (b) and for sample B in (d). The vertical dashed black line represents the EEZ wavelength. The maximum pump intensity used for extracting the $\chi^{(3)}$ values was 21.5 GW/cm$^2$ for sample A and 14.6 GW/cm$^2$ for sample B. \label{fig4}}
\end{center}
\end{figure}

Two sets of measurements for pump intensities of 20 GW/cm$^2$ and 10 GW/cm$^2$ are shown in Fig.~\ref{fig2}(a) and (b) for samples A and B, respectively. We  note that there is a clear step-like increase in the normalised $\Delta R/R_{\rm lin}$ as the probe  is tuned across the EEZ wavelength. We didn't observe any variation in the transmissivity ($\Delta T=0$) within the noise limits of our measurements (a few percent).
 \\
As we will show in the following, different effects contribute to the change in reflectivity $\Delta R/R_{\rm lin}$, including the dispersion of the $\chi^{(3)}$ and the transition through the EEZ wavelength. A more thorough analysis of the variation of the permittivity with the pump intensity and probe wavelength is therefore required to unveil the underlying processes.
We thus extend the same method followed to extract the linear permittivity $\varepsilon$ from $R_{\rm lin}$ and $T_{\rm lin}$, to also extract $\varepsilon$ in the presence of the pump: the values of the reflectivity and transmissivity in the nonlinear (pumped) case allow to retrieve the nonlinear (pumped) value of the permittivity. In particular, here we are interested in the behaviour just above the EEZ wavelength: as can be seen in Fig.~\ref{fig1}, here the unpumped $\varepsilon '$ is negative. In the presence of a positive and sufficiently large increase in $\varepsilon '$ due to the optical pump we may predict that the permittivity will transition from below to above zero. Figure~\ref{fig3a} shows $\varepsilon '$ as a function of pump intensity for sample A  and sample B, measured at 890 nm and 825 nm, respectively (in both cases, 5 nm above the EEZ wavelength). As can be seen, in both cases the permittivity transitions from negative to positive, thus indicating a transition of the medium from metallic to dielectric. The total variation $\Delta\varepsilon '\sim0.05$ is of the same order of the absolute value of the permittivity itself, implying a relatively large bandwidth of $\sim10$ nm over which the optically-induced metal-dielectric transition occurs for the maximum pump power (limited by material damage). In the inset (a) to Fig.~\ref{fig3a} we also show the corresponding imaginary parts of the permittivity that also increase with pump intensity. Inset (b) shows $\varepsilon '$ as a function of the relative pump-probe delay measured on sample A for a pump power of 17 GW/cm$^2$: the rise time is of the order of the 100 fs pump pulse duration (followed by a decay time of a few ps that is typical for Ag). We see that by tuning the delay, it is possibly to tune the precise value of $\varepsilon '$, crossing from metal to dielectric and back and again.
\\
A notable feature of this data is the clear linear dependence of both $\varepsilon '$ and $\varepsilon ''$ with pump power (in disagreement for example with the theoretical predictions of Husakou et al.~\cite{husakou}): the dashed lines in Fig.~\ref{fig3a} represent linear fits to the data, which are seen to pass through the value measured in the absence of the pump (and reported in Fig.~\ref{fig1}, as expected). We also note that this linear behaviour was observed over a wide range of wavelengths (700 nm to 1000 nm, data not shown).  This feature is remarkable as it allows us to relate the behaviour of the MM and the transition from metal to dielectric in terms of a standard third order nonlinear susceptibility,  $\chi^{(3)}$ \cite{boyd}. Indeed, because of this linear behaviour we can extarct the nonlinear susceptibility tensor element as (see Methods for details):
\begin{equation}\label{chi3}
\chi^{(3)}(\opr,\op) = \frac{n_p\veps_0 c}{3}\frac{\partial \veps(\opr,I_p)}{\partial I_p},
\end{equation}
where $\opr$ and $\op$ are the probe and pump frequencies, respectively, and $n_p$ is the real part of the medium refractive index at the pump frequency.\\
The $\chi^{(3)}$ values are therefore determined by the intensity gradient of $\veps$ retrieved form the reflectivity and transmissivity measurements at different pump intensities. We note that the linearity observed in the variation of $\varepsilon$ with pump intensity (as shown in Fig.~\ref{fig3a}) implies that Eq.~\eqref{chi3} is consistent, as it provides us with $\chi^{(3)}$ values that are constants (do not depend on $I_p$). We remark that this method differs from the simple retrieval of the permittivity in the pumped case, as it exploits the found linear behaviour of $\varepsilon$ versus intensity to interpret the nonlinear mechanism in term of a third-order nonlinearity, also allowing to extract the complex value of the $\chi^{(3)}$ tensor at different pump and probe wavelengths.\\
This derivation neglects the variation of the pump intensity inside the sample along the propagation direction (due to absorption).
Averaging the intensity over the sample thickness would lead to a small correction factor $\sim2$ for the values of $\chi^{(3)}$, thus here we chose the simplified formulation.\\
We use Eq.~\eqref{chi3}, applied separately to the real and imaginary part of $\varepsilon$ to plot the real and imaginary parts of $\chi^{(3)}$. Figures~\ref{fig4}(a) and (c) show the real (solid blue line) and imaginary (dashed red line) third order nonlinear coefficients $\chi^{(3)}_r$  and $\chi^{(3)}_i$  for samples A  and B, respectively. Figures (b) and (d) show the same data but plotted as the absolute value $|\chi^{(3)}|$ (solid blue line) and phase, $\phi$ (dashed red line). The notable feature of these results is that as the probe wavelength crosses from the dielectric-like region ($\veps '>0$) to the metallic-like region ($\veps '<0$), the $|\chi^{(3)}|$ changes in nature from predominantly real to predominantly imaginary.
\section*{Discussion}
The maximum measured $|\chi^{(3)}|$ is of the same order or slightly larger than that measured by other means in bulk silver $|\chi^{(3)}|=2.8\,10^{-19}$ m$^2$/V$^2$ \cite{bloembergen,boyd}. However, it is interesting to note that whereas silver cannot be used in transmission in samples thicker than $\sim40$ nm (all light is reflected from the surface at these thicknesses), the metamaterial is significantly thicker and more than two orders thicker than the total thickness of silver present in the sample. This therefore provides a much longer effective interaction length for light with the high nonlinearity of the metal, as compared to the bare material. Note that the increased transmission is a known result for alternating layers of dielectric and metals, and it is for example exploited in one-dimensional metal-dielectric photonic crystal \cite{boyd_PC_theo,boyd_PC}.\\
We also underline that the large imaginary component of the $\chi^{(3)}$ tensor measured in the ``metallic'' wavelength region does not imply larger nonlinear losses. Indeed, third order nonlinearities lead to nonlinear phase shifts and nonlinear absorption that are determined by the real and imaginary part of nonlinear refractive index, respectively (and not of the susceptibility). These are given by (in the degenerate case) \cite{boyd2}: $n_{2r}=3/(4\varepsilon_0 c){[n_r\chi^{(3)}_r+n_i\chi^{(3)}_i]/D}$ and  $n_{2i}=3/(4\varepsilon_0 c){[n_r\chi^{(3)}_r-n_i\chi^{(3)}_i]/D}$ where $D=n_r(n_r^2+n_i^2)$. The imaginary part of $n_2$ is usually associated to what is known as the nonlinear absorption coefficient, $\beta_2=4\pi n_{2i}/\lambda$, where $\lambda$ is the vacuum wavelength. We underline that $n_{2r}$ and $\beta_2$ depend respectively, on the sums and differences of the real and imaginary parts of  of the $\chi^{(3)}$ tensor. This implies for example that a large $\chi^{(3)}_i$ will enhance Kerr effects such as phase modulation that are associated to the $n_{2r}$ coefficient whilst simultaneously minimising nonlinear absorption (the data in Fig.~\ref{fig4} shows that actually $\beta_2\sim0$ close to the EEZ wavelength).\\
In conclusion, EEZ metamaterials allow to tailor and access novel optical propagation regimes. The interplay between the linear and nonlinear, real and imaginary propagation constants leads to metamaterials that exhibit nonlinearities with amplitudes similar to those of metals yet spread over material thicknesses two orders of magnitude larger and with substantially reduced losses. We have also shown that it is possible to optically induce a metal-dielectric transition over a wide bandwidth that can support $\sim100$ fs laser pulses. 
Further optimisation by e.g. reducing losses (and hence the linear refractive index) by introducing a gain medium \cite{Rizza_11} may enable new forms of efficient switches for light and even a platform for non-perturbative nonlinear optics at low light intensities. 

\section*{Methods}

{\bf Fabrication:} The ENZ metamaterial was fabricated on 1 mm thick slides of SiO$_2$. The substrate was thoroughly cleaned in ultrasonic assisted baths of Acetone and Isopropanol (5 mins each) and blown dried with N$_2$ flow. The metallo-dielectric stack was deposited using a EDWARD auto $306$ electron beam evaporator, with a base pressure below $3\times10^{-6}$ bar. The deposition rate was kept below $0.1$ nm/s for the metals and below $0.3$ nm/s for the SiO$_2$ to grant uniformity. The thickness of the seeding Ge layer (0.7 nm) was chosen after a thorough optimisation process, to minimise the losses of the thinnest achievable layer of Silver (5 nm). An excess of 50 independent evaporations was performed to determine the final values. The development process was complemented by a combination of SEM, STEM and AFM measurements on sacrificial test samples.\\

\noindent 
{\bf Linear characterisation:} To characterise the linear properties of the MM we built a simple setup for the measurement of the reflection and the transmission. An Ocean Optics HL2000 Halogen Lamp covering the visible-NIR spectrum was collimated with a  telescope to a beam radius of $\simeq5 mm$, with controlled polarisation. A thin film beam splitter was used to separate the reflected light from the incident one. The two beams were then focused on the tip of multimode fibres and analysed with two Ocean Optics spectrum analysers. Light was impinging on the sample on the side of the multilayer. The reference for the reflection was a silver mirror, for the transmission we recorded the collected light without any sample. The raw data were then used to retrieve the effective complex permittivity of the multilayer, using a standard least mean square fitting procedure: the experimental reflected and transmitted value vs $\lambda$ where compared with those calculated with a transmission matrix simulation, using a test value for the permittivity. Using the same method we also characterised the linear permittivity of SiO$_2$ and that of a single Ge/Ag bilayer, which were then used to calculate the effective index permittivity of the homogenised stack.\\

\noindent
{\bf Nonlinear characterisation}
The material polarisation at the probe wavelength is given by 
\begin{equation}
P(\opr)= \veps_0 \left[\veps(\opr,I_p)-1\right] E(\opr)\,,
\end{equation}
 where $\opr$ and $\op$ are the probe and pump frequencies, respectively, $E(\opr)$ is the probe electric field, and the relative nonlinear permittivity is defined as
\begin{equation}
\veps(\opr,I_p) =1+\chi^{(1)}(\opr)+6\chi^{(3)}(\opr,\op)\frac{I_p}{2 n_p\veps_0 c}.
\end{equation}
being $n_p$  the real part of the medium refractive index at the pump frequency. It is thus clear that a linear scaling of the permittivity with the pump intensity allows an interpretation in terms of a third-order nonlinearity, where the the $\chi^{(3)}$ tensor can be retrieved by deriving the relative permittivity $\veps(\opr,I_p)$ with respect to $I_p$:
\begin{equation}\label{chi3_methods}
\chi^{(3)}(\opr,\op) = \frac{n_p\veps_0 c}{3}\frac{\partial \veps(\opr,I_p)}{\partial I_p},
\end{equation}\\
The nonlinear (pumped) permittivity $\veps(\opr,I_p)$ has been evaluated from the reflectivity and transmissivity measurements in presence of the pump, for different pump intensities and probe wavelength. Starting from the linear $\veps(\opr,0)$ and exploiting a transfer matrix approach similar to the one used for the linear characterisation, we found the value $\veps(\opr,I_p)$ resulting in a $\Delta R/R_{\rm lin}$ and $\Delta T/T_{\rm lin}$ that best matched the experimental data at the given $I_p$. The dual condition on reflectivity and transmissivity, allows to retrieve both the real and imaginary part of $\veps(\opr,I_p)$.
We then used Eq.~\ref{chi3_methods} (same as Eq.~\ref{chi3} in the main text) to calculate $\chi^{(3)}(\opr,\op)$.

\section*{Acknowledgements}

This work was supported by the EPSRC grant EP/ J004200/1. D.F. acknowledges financial support from the European Research Council under the European Union Seventh Framework Programme (FP/2007-2013)/ERC GA 306559 and EPSRC (UK, Grant No. EP/J00443X/1). L.C. and M.C. acknowledge the support from the People Programme (Marie Curie Actions) of the European Union's FP7 Programme THREEPLE (GA 627478) and KOHERENT (GA 299522). A.C. and C.R. acknowledge support from U.S. Army International Technology Center Atlantic for financial support (Grant No. W911NF-14-1-0315). The authors acknowledge discussions with M. Scalora.

\section*{Author contributions statement}

R.K.K. performed experiments with assistance from T.R. and M.C., M.P. fabricated and characterised linearly the samples, L.C. and M.P. performed data and theoretical analysis, C.R. and A.C. provided theoretical support, D.F. and A.D.F. directed the project and wrote the manuscript. All authors contributed to scientific discussions and to the manuscript.

\section*{Additional information}

\textbf{Competing financial interests:} The authors declare no competing financial interests. 


\begin{thebibliography}{36}
\expandafter\ifx\csname url\endcsname\relax
  \def\url#1{\texttt{#1}}\fi
\expandafter\ifx\csname urlprefix\endcsname\relax\def\urlprefix{URL }\fi
\providecommand{\bibinfo}[2]{#2}
\providecommand{\eprint}[2][]{\url{#2}}

\bibitem{alu1}
\bibinfo{author}{Al\`u, A.}, \bibinfo{author}{Silveirinha, M.~G.},
  \bibinfo{author}{Salandrino, A.} \& \bibinfo{author}{Engheta, N.}
\newblock \bibinfo{title}{Epsilon-near-zero metamaterials and electromagnetic
  sources: Tailoring the radiation phase pattern}.
\newblock \emph{\bibinfo{journal}{Phys. Rev. B}} \textbf{\bibinfo{volume}{75}},
  \bibinfo{pages}{155410} (\bibinfo{year}{2007}).

\bibitem{alu2}
\bibinfo{author}{Edwards, B.}, \bibinfo{author}{Al\`u, A.},
  \bibinfo{author}{Young, M.~E.}, \bibinfo{author}{Silveirinha, M.} \&
  \bibinfo{author}{Engheta, N.}
\newblock \bibinfo{title}{Experimental verification of epsilon-near-zero
  metamaterial coupling and energy squeezing using a microwave waveguide}.
\newblock \emph{\bibinfo{journal}{Phys. Rev. Lett.}}
  \textbf{\bibinfo{volume}{100}}, \bibinfo{pages}{033903}
  (\bibinfo{year}{2008}).

\bibitem{eps0_1}
\bibinfo{author}{Ziolkowski, R.~W.}
\newblock \bibinfo{title}{{Propagation in and scattering from a matched
  metamaterial having a zero index of refraction.}}
\newblock \emph{\bibinfo{journal}{Phys. Rev. E}} \textbf{\bibinfo{volume}{70}},
  \bibinfo{pages}{046608} (\bibinfo{year}{2004}).

\bibitem{eps0_2}
\bibinfo{author}{Zhou, H.} \emph{et~al.}
\newblock \bibinfo{title}{{A high-directive patch antenna based on
  all-dielectric near-zero-index metamaterial superstrates}}.
\newblock \emph{\bibinfo{journal}{J. Electromagn. Waves Appl.}}
  \textbf{\bibinfo{volume}{24}}, \bibinfo{pages}{1387–1396}
  (\bibinfo{year}{2010}).

\bibitem{eps0_3}
\bibinfo{author}{Yang, J.}, \bibinfo{author}{Huang, M.} \&
  \bibinfo{author}{Peng, J.}
\newblock \bibinfo{title}{{Directive emission obtained by Mu and
  epsilon-near-zero metamaterials.}}
\newblock \emph{\bibinfo{journal}{Radioengineering}}
  \textbf{\bibinfo{volume}{18}}, \bibinfo{pages}{124–128}
  (\bibinfo{year}{2009}).

\bibitem{eps0_4}
\bibinfo{author}{Soric, J.~C.}, \bibinfo{author}{Engheta, N.},
  \bibinfo{author}{Maci, S.} \& \bibinfo{author}{Al{\`{u}}, A.}
\newblock \bibinfo{title}{{Omnidirectional metamaterial antennas based on
  e-near-zero channel matching.}}
\newblock \emph{\bibinfo{journal}{IEEE Trans. Antennas Propag.}}
  \textbf{\bibinfo{volume}{61}}, \bibinfo{pages}{33–44}
  (\bibinfo{year}{2013}).

\bibitem{eps0_5}
\bibinfo{author}{Jin, Y.} \& \bibinfo{author}{He, S.}
\newblock \bibinfo{title}{{Enhancing and suppressing radiation with some
  permeability-near-zero structures.}}
\newblock \emph{\bibinfo{journal}{Opt. Express}} \textbf{\bibinfo{volume}{18}},
  \bibinfo{pages}{16587–16593} (\bibinfo{year}{2010}).

\bibitem{eps0_6}
\bibinfo{author}{Silveirinha, M.} \& \bibinfo{author}{Engheta, N.}
\newblock \bibinfo{title}{{Tunneling of electromagnetic energy through
  subwavelength channels and bends using {$\varepsilon$}-near-zero materials.}}
\newblock \emph{\bibinfo{journal}{Phys. Rev. Lett.}}
  \textbf{\bibinfo{volume}{97}}, \bibinfo{pages}{157403}
  (\bibinfo{year}{2006}).

\bibitem{eps0_7}
\bibinfo{author}{Silveirinha, M.} \& \bibinfo{author}{Engheta, N.}
\newblock \bibinfo{title}{{Design of matched zero-index metamaterials using
  nonmagnetic inclusions in epsilon-near-zero media.}}
\newblock \emph{\bibinfo{journal}{Phys. Rev. B}} \textbf{\bibinfo{volume}{75}},
  \bibinfo{pages}{075119} (\bibinfo{year}{2007}).

\bibitem{eps0_8}
\bibinfo{author}{Pan, Y.} \& \bibinfo{author}{Xu, S.}
\newblock \bibinfo{title}{{ Energy tunnelling through an ultrasmall
  epsilon-near-zero channel in circular waveguide.}}
\newblock \emph{\bibinfo{journal}{IET Microw. Antennas Propag.}}
  \textbf{\bibinfo{volume}{3}}, \bibinfo{pages}{821–825}
  (\bibinfo{year}{2009}).

\bibitem{eps0_9}
\bibinfo{author}{Al{\`{u}}, A.} \& \bibinfo{author}{Engheta, N.}
\newblock \bibinfo{title}{{Coaxial-to-waveguide matching with
  $\varepsilon$-near-zero ultranarrow channels and bends.}}
\newblock \emph{\bibinfo{journal}{IEEE Trans. Antennas Propag.}}
  \textbf{\bibinfo{volume}{58}}, \bibinfo{pages}{328–339}
  (\bibinfo{year}{2010}).

\bibitem{eps0_11}
\bibinfo{author}{Nguyen, V.~C.}, \bibinfo{author}{Chen, L.} \&
  \bibinfo{author}{Halterman, K.}
\newblock \bibinfo{title}{{Total transmission and total reflection by zero
  index metamaterials with defects.}}
\newblock \emph{\bibinfo{journal}{Phys. Rev. Lett.}}
  \textbf{\bibinfo{volume}{105}}, \bibinfo{pages}{233908}
  (\bibinfo{year}{2010}).

\bibitem{eps0_12}
\bibinfo{author}{Feng, S.}
\newblock \bibinfo{title}{{ Loss-induced omnidirectional bending to the normal
  in {$\varepsilon$}-near-zero metamaterials.}}
\newblock \emph{\bibinfo{journal}{Phys. Rev. Lett.}}
  \textbf{\bibinfo{volume}{108}}, \bibinfo{pages}{193904}
  (\bibinfo{year}{2012}).

\bibitem{eps0_15}
\bibinfo{author}{Fleury, R.} \& \bibinfo{author}{Al{\`{u}}, A.}
\newblock \bibinfo{title}{{Enhanced superradiance in epsilon-near-zero
  plasmonic channels.}}
\newblock \emph{\bibinfo{journal}{Phys. Rev. B}} \textbf{\bibinfo{volume}{87}},
  \bibinfo{pages}{201101} (\bibinfo{year}{2013}).

\bibitem{eps0_14}
\bibinfo{author}{Maas, R.}, \bibinfo{author}{Parsons, J.},
  \bibinfo{author}{Engheta, N.} \& \bibinfo{author}{Polman, A.}
\newblock \bibinfo{title}{{ Experimental realization of an epsilon-near-zero
  metamaterial at visible wavelengths.}}
\newblock \emph{\bibinfo{journal}{Nature Photon.}}
  \textbf{\bibinfo{volume}{7}}, \bibinfo{pages}{907–912}
  (\bibinfo{year}{2013}).

\bibitem{eps0_13}
\bibinfo{author}{Argyropoulos, C.}, \bibinfo{author}{Chen, P.},
  \bibinfo{author}{D’Aguanno, N., G.~Engheta} \& \bibinfo{author}{Al{\`{u}},
  A.}
\newblock \bibinfo{title}{{ Boosting optical nonlinearities in
  {$\varepsilon$}-near-zero plasmonic channels.}}
\newblock \emph{\bibinfo{journal}{Phys. Rev. B}}
  \textbf{\bibinfo{volume}{185}}, \bibinfo{pages}{045129}
  (\bibinfo{year}{2012}).

\bibitem{ciattoni:2010}
\bibinfo{author}{Ciattoni, A.}, \bibinfo{author}{Rizza, C.} \&
  \bibinfo{author}{Palange, E.}
\newblock \bibinfo{title}{Extreme nonlinear electrodynamics in metamaterials
  with very small linear dielectric permittivity}.
\newblock \emph{\bibinfo{journal}{Phys. Rev. A}} \textbf{\bibinfo{volume}{81}},
  \bibinfo{pages}{043839} (\bibinfo{year}{2010}).

\bibitem{ciattoni2}
\bibinfo{author}{Vincenti, M.~A.}, \bibinfo{author}{de~Ceglia, D.},
  \bibinfo{author}{Ciattoni, A.} \& \bibinfo{author}{Scalora, M.}
\newblock \bibinfo{title}{Singularity-driven second- and third-harmonic
  generation at $\epsilon$-near-zero crossing points}.
\newblock \emph{\bibinfo{journal}{Phys. Rev. A}} \textbf{\bibinfo{volume}{84}},
  \bibinfo{pages}{063826} (\bibinfo{year}{2011}).

\bibitem{ciattoni3}
\bibinfo{author}{Rizza, C.}, \bibinfo{author}{Ciattoni, A.} \&
  \bibinfo{author}{Palange, E.}
\newblock \bibinfo{title}{Two-peaked and flat-top perfect bright solitons in
  nonlinear metamaterials with epsilon near zero}.
\newblock \emph{\bibinfo{journal}{Phys. Rev. A}} \textbf{\bibinfo{volume}{83}},
  \bibinfo{pages}{053805} (\bibinfo{year}{2011}).

\bibitem{Ciattoni:10}
\bibinfo{author}{Ciattoni, A.}, \bibinfo{author}{Rizza, C.} \&
  \bibinfo{author}{Palange, E.}
\newblock \bibinfo{title}{Transverse power flow reversing of guided waves in
  extreme nonlinear metamaterials}.
\newblock \emph{\bibinfo{journal}{Opt. Express}} \textbf{\bibinfo{volume}{18}},
  \bibinfo{pages}{11911--11916} (\bibinfo{year}{2010}).
\newblock

\bibitem{zayats}
\bibinfo{author}{Kauranen, M.} \& \bibinfo{author}{Zayats, A.}
\newblock \bibinfo{title}{Nonlinear plasmonics}.
\newblock \emph{\bibinfo{journal}{Nature Photon.}} \textbf{\bibinfo{volume}{6}},
  \bibinfo{pages}{737} (\bibinfo{year}{2012}).

\bibitem{Rizza_11}
\bibinfo{author}{Rizza, C.}, \bibinfo{author}{Di~Falco, A.} \&
  \bibinfo{author}{Ciattoni, A.}
\newblock \bibinfo{title}{{Gain assisted nanocomposite multilayers with near
  zero permittivity modulus at visible frequencies}}.
\newblock \emph{\bibinfo{journal}{App. Phys. Lett.}}
  \textbf{\bibinfo{volume}{99}}, \bibinfo{pages}{221107}
  (\bibinfo{year}{2011}).

\bibitem{Suchowski:2013fa}
\bibinfo{author}{Suchowski, H.} \emph{et~al.}
\newblock \bibinfo{title}{{Phase Mismatch-Free Nonlinear Propagation in Optical
  Zero-Index Materials}}.
\newblock \emph{\bibinfo{journal}{Science}} \textbf{\bibinfo{volume}{342}},
  \bibinfo{pages}{1223--1226} (\bibinfo{year}{2013}).

\bibitem{husakou}
\bibinfo{author}{Husakou, A.} \& \bibinfo{author}{Hermann, J.}
\newblock \bibinfo{title}{Steplike transmission of light through a
  metal-dielectric multilayer structure due to an intensity-dependent sign of
  the effective dielectric constant}.
\newblock \emph{\bibinfo{journal}{Phys. Rev. Lett.}}
  \textbf{\bibinfo{volume}{99}}, \bibinfo{pages}{127402}
  (\bibinfo{year}{2007}).

\bibitem{ITO}
\bibinfo{author}{Capretti, A.}, \bibinfo{author}{Y., W.} \&
  \bibinfo{author}{L., E. N. D.~N.}
\newblock \bibinfo{title}{{Enhanced third-harmonic generation in Si-compatible
  epsilon-near-zero indium tin oxide nanolayers.}}
\newblock \emph{\bibinfo{journal}{Opt. Lett.}} \textbf{\bibinfo{volume}{40}},
  \bibinfo{pages}{1500--1503} (\bibinfo{year}{2015}).

\bibitem{VO2}
\bibinfo{author}{Lysenko, S.}, \bibinfo{author}{Rua, A.},
  \bibinfo{author}{Fernandez, F.} \& \bibinfo{author}{Liu, H.}
\newblock \bibinfo{title}{Optical nonlinearity and structural dynamics of vo2
  films}.
\newblock \emph{\bibinfo{journal}{J. Appl. Phys.}}
  \textbf{\bibinfo{volume}{105}}, \bibinfo{pages}{043502}
  (\bibinfo{year}{2009}).

\bibitem{Krausz_nature1}
\bibinfo{author}{Schultze, M.} \emph{et~al.}
\newblock \bibinfo{title}{Controlling dielectrics with the electric field of
  light}.
\newblock \emph{\bibinfo{journal}{Nature}} \textbf{\bibinfo{volume}{493}},
  \bibinfo{pages}{75} (\bibinfo{year}{2013}).

\bibitem{Krausz_nature2}
\bibinfo{author}{Schiffrin, A.} \emph{et~al.}
\newblock \bibinfo{title}{Optical-field-induced current in dielectrics}.
\newblock \emph{\bibinfo{journal}{Nature}} \textbf{\bibinfo{volume}{493}},
  \bibinfo{pages}{70} (\bibinfo{year}{2013}).

\bibitem{Shalaev_10}
\bibinfo{author}{Chen, W.}, \bibinfo{author}{Thoreson, M.~D.},
  \bibinfo{author}{Ishii, S.}, \bibinfo{author}{Kildishev, A.~V.} \&
  \bibinfo{author}{Shalaev, V.~M.}
\newblock \bibinfo{title}{{Ultra-thin ultra-smooth and low-loss silver films on
  a germanium wetting layer}}.
\newblock \emph{\bibinfo{journal}{Opt. Express}}
  \textbf{\bibinfo{volume}{18}}, \bibinfo{pages}{5124--5134}
  (\bibinfo{year}{2010}).

\bibitem{sipe:1992}
\bibinfo{author}{Sipe, J.} \& \bibinfo{author}{Boyd, R.}
\newblock \bibinfo{title}{Nonlinear susceptibility of composite optical
  materials in the maxwell-garnett model}.
\newblock \emph{\bibinfo{journal}{Phys. Rev. A}} \textbf{\bibinfo{volume}{46}},
  \bibinfo{pages}{1614} (\bibinfo{year}{1992}).

\bibitem{maxwell_garnett}
\bibinfo{author}{Garnett, J.~M.}
\newblock \bibinfo{title}{Enhanced nonlinear optical response of composite
  materials}.
\newblock \emph{\bibinfo{journal}{Philos. Trans. R. Soc. London}}
  \textbf{\bibinfo{volume}{203}}, \bibinfo{pages}{385} (\bibinfo{year}{1904}).

\bibitem{boyd}
\bibinfo{author}{Boyd, R.}
\newblock \emph{\bibinfo{title}{Nonlinear Optics}}
  (\bibinfo{publisher}{Academic Press}, \bibinfo{address}{New York},
  \bibinfo{year}{2008}).

\bibitem{bloembergen}
\bibinfo{author}{Bloembergen, N.}, \bibinfo{author}{Burns, W.} \&
  \bibinfo{author}{Matsuoka, M.}
\newblock \bibinfo{title}{Reflected third harmonic generated by picosecond
  laser pulses}.
\newblock \emph{\bibinfo{journal}{Opt. Commun.}} \textbf{\bibinfo{volume}{1}},
  \bibinfo{pages}{195--198} (\bibinfo{year}{1969}).

\bibitem{boyd_PC_theo}
\bibinfo{author}{Bennink, R.~S.}, \bibinfo{author}{Yoon, Y.-K.},
  \bibinfo{author}{Boyd, R.~W.} \& \bibinfo{author}{Sipe, J.~E.}
\newblock \bibinfo{title}{Accessing the optical nonlinearity of metals with
  metal--dielectric photonic bandgap structures}.
\newblock \emph{\bibinfo{journal}{Opt. Lett.}} \textbf{\bibinfo{volume}{24}},
  \bibinfo{pages}{1416--1418} (\bibinfo{year}{1999}).

\bibitem{boyd_PC}
\bibinfo{author}{Lepeshkin, N.}, \bibinfo{author}{Schweinsberg, A.},
  \bibinfo{author}{Piredda, G.}, \bibinfo{author}{Bennink, R.} \&
  \bibinfo{author}{Boyd, R.}
\newblock \bibinfo{title}{Enhanced nonlinear optical response of
  one-dimensional metal-dielectric photonic crystals}.
\newblock \emph{\bibinfo{journal}{Phys. Rev. Lett.}}
  \textbf{\bibinfo{volume}{93}}, \bibinfo{pages}{123902}
  (\bibinfo{year}{2004}).

\bibitem{boyd2}
\bibinfo{author}{Smith, D.~D.} \emph{et~al.}
\newblock \bibinfo{title}{{Z}-scan measurement of the nonlinear absorption of a
  thin gold film}.
\newblock \emph{\bibinfo{journal}{J. Appl. Phys.}}
  \textbf{\bibinfo{volume}{86}}, \bibinfo{pages}{6200} (\bibinfo{year}{1999}).

\end{thebibliography}
\end{document}